\documentclass[12pt,letterpaper]{article}
\usepackage[margin=1in]{geometry}
\usepackage[english]{babel}
\usepackage[utf8x]{inputenc}
\usepackage[T1]{fontenc}

\usepackage{amsmath}
\usepackage{amssymb}
\usepackage{graphicx}
\usepackage{adjustbox}
\usepackage[colorinlistoftodos]{todonotes}
\usepackage[colorlinks=true, allcolors=blue]{hyperref}
\usepackage{subfigure}
\usepackage[toc,page]{appendix} 
\usepackage{wrapfig}
\usepackage{cite}

\newcommand{\ANLHEP}{HEP Division, Argonne National Laboratory, Lemont, IL 60439, USA}
\newcommand{\APC}{Laboratoire Astroparticule et Cosmologie (APC), CNRS/IN2P3, Universit\'e Paris Diderot, 10, rue Alice Domon et Léonie Duquet, 75205 Paris Cedex 13, France}

\newcommand{\BenGurion}{Department of Physics, Ben-Gurion University, Be'er Sheva 84105, Israel}
\newcommand{\BNL}{Brookhaven National Laboratory, Upton, NY 11973}
\newcommand{\Brown}{Brown University, Providence, RI 02912}

\newcommand{\Caltech}{California Institute of Technology, Pasadena, CA 91125}
\newcommand{\Cardiff}{School of Physics and Astronomy, Cardiff University, The Parade, Cardiff, CF24 3AA, UK}

\newcommand{\CCA}{Center for Computational Astrophysics, 162 5th Ave, 10010, New York, NY, USA}
\newcommand{\CPPM}{Aix Marseille Univ, CNRS/IN2P3, CPPM, Marseille, France}
\newcommand{\CEADAP}{D\'epartement d’Astrophysique, CEA Saclay DSM/Irfu, 91191 Gif-sur-Yvette, France}

\newcommand{\CFT}{Center for Theoretical Physics, Polish Academy of Sciences, al. Lotnik\'{o}w 32/46, 02-668, Warsaw, Poland}

\newcommand{\CITA}{Canadian Institute for Theoretical Astrophysics, University of Toronto, Toronto, ON M5S 3H8, Canada}

\newcommand{\CMUCosmo}{Department 
of Physics, McWilliams Center for Cosmology, Carnegie Mellon University}

\newcommand{\Cornell}{Cornell University, Ithaca, NY 14853}

\newcommand{\daa}{Department of Astronomy and Astrophysics, University of Toronto, ON, M5S3H4}
\newcommand{\damtp}{DAMTP, Centre for Mathematical Sciences, Wilberforce Road, Cambridge, UK, CB3 0WA}

\newcommand{\dunlap}{Dunlap Institute for Astronomy and Astrophysics, University of Toronto, ON, M5S3H4}
\newcommand{\Durham}{Department of Physics, Lower Mountjoy, South Rd, Durham DH1 3LE, United Kingdom}
\newcommand{\ED}{University of Edinburgh, EH8 9YL Edinburgh, United Kingdom}
\newcommand{\EPFL}{Laboratory of Astrophysics, EPFL, 1290 Versoix, Switzerland}

\newcommand{\FNAL}{Fermi National Accelerator Laboratory, Batavia, IL 60510}
\newcommand{\FQAUB}{Dept. de F\' isica Qu\` antica i Astrof\' isica, Universitat de Barcelona, Mart\' i i Franqu\` es 1, E08028 Barcelona, Spain}

\newcommand{\GSFC}{Goddard Space Flight Center, Greenbelt, MD 20771 USA}

\newcommand{\HarvardPhys}{Department of Physics, Harvard University, Cambridge, MA 02138, USA}

\newcommand{\IAP}{Institut d'Astrophysique de Paris (IAP), CNRS \& Sorbonne University, Paris, France}
\newcommand{\IAS}{Institute for Advanced Study, Princeton, NJ 08540}

\newcommand{\ICC}{ICC, University of Barcelona, IEEC-UB, Mart\' i i Franqu\` es, 1, E08028 Barcelona, Spain}
\newcommand{\ICE}{Institute of Space Sciences (ICE, CSIC), Campus UAB, Carrer de Can Magrans, s/n, 08193 Barcelona, Spain}

\newcommand{\ICTP}{International Centre for Theoretical Physics, Strada Costiera, 11, I-34151 Trieste, Italy}
\newcommand{\IFAE}{Institut de Fisica d’Altes Energies, The Barcelona Institute of Science and Technology, Campus UAB, 08193 Bellaterra (Barcelona), Spain}
\newcommand{\IFPU}{IFPU - Institute for Fundamental Physics of the Universe, Via Beirut 2, 34014 Trieste, Italy}
\newcommand{\IFT}{Instituto de Fisica Teorica UAM/CSIC, Universidad Autonoma de Madrid, 28049 Madrid, Spain}
\newcommand{\IFUNAM}{IFUNAM - Instituto de F\'isica, Universidad Nacional Aut\'onoma de M\'etico, 04510 CDMX, M\'exico}

\newcommand{\INAFOATs}{INAF - Osservatorio Astronomico di Trieste, Via G.B. Tiepolo 11, 34143 Trieste, Italy}

\newcommand{\INFN}{INFN – National Institute for Nuclear Physics, Via Valerio 2, I-34127 Trieste, Italy}
\newcommand{\INFNFerrara}{Istituto Nazionale di Fisica Nucleare, Sezione di Ferrara, Polo Scientifico e Tecnologico, Edificio C, Via Saragat 1, I-44122 Ferrara, Italy}

\newcommand{\INFNRM}{Istituto Nazionale di Fisica Nucleare, Sezione di Roma, 00185 Roma, Italy}

\newcommand{\ioa}{Institute of Astronomy, University of Cambridge,Cambridge CB3 0HA, UK}

\newcommand{\IRFU}{IRFU, CEA, Universit\'e Paris-Saclay, F-91191 Gif-sur-Yvette, France}
\newcommand{\IUCAA}{The Inter-University Centre for Astronomy and Astrophysics, Pune, 411007, India}

\newcommand{\JHU}{Johns Hopkins University, Baltimore, MD 21218}

\newcommand{\JPL}{Jet Propulsion Laboratory, California Institute of Technology, Pasadena, CA, USA}
\newcommand{\KASSI}{Korea Astronomy and Space Science Institute, Daejeon 34055, Korea}
\newcommand{\kavli}{Kavli Institute for Cosmology, Cambridge, UK, CB3 0HA}
\newcommand{\KICP}{Kavli Institute for Cosmological Physics, Chicago, IL 60637}
\newcommand{\KIPAC}{Kavli Institute for Particle Astrophysics and Cosmology, Stanford 94305}

\newcommand{\KSU}{Kansas State University, Manhattan, KS 66506}

\newcommand{\Lagrange}{Sorbonne Universit\'e, Institut Lagrange de Paris (ILP), 98 bis boulevard Arago, F-75014 Paris, France}

\newcommand{\LBL}{Lawrence Berkeley National Laboratory, Berkeley, CA 94720}

\newcommand{\LLNL}{Lawrence Livermore National Laboratory, Livermore, CA, 94550}
\newcommand{\McGill}{McGill University, Montreal, QC H3A 2T8, Canada}
\newcommand{\Melbourne}{School of Physics, The University of Melbourne, Parkville, VIC 3010, Australia}

\newcommand{\MPE}{Max-Planck-Institut f\"{u}r extraterrestrische Physik (MPE), Giessenbachstrasse 1, D-85748 Garching bei M\"unchen, Germany}

\newcommand{\NAOC}{National Astronomical Observatories, Chinese Academy of Sciences, PR China}

\newcommand{\OSU}{The Ohio State University, Columbus, OH 43212}
\newcommand{\OU}{Department of Physics and Astronomy, Ohio University, Clippinger Labs, Athens, OH 45701, USA}

\newcommand{\Oxford}{The University of Oxford, Oxford OX1 3RH, UK}

\newcommand{\PI}{Perimeter Institute, Waterloo, Ontario N2L 2Y5, Canada}

\newcommand{\Port}{Institute of Cosmology \& Gravitation, University of Portsmouth, Dennis Sciama Building, Burnaby Road, Portsmouth PO1 3FX, UK}
\newcommand{\Princeton}{Princeton University, Princeton, NJ 08544}

\newcommand{\Queensland}{The University of Queensland, School of Mathematics and Physics, QLD 4072, Australia}
\newcommand{\QMUL}{Queen Mary University of London, Mile End Road, London E1 4NS, United Kingdom}

\newcommand{\Rice}{Department of Physics \& Astronomy, Rice University, Houston, Texas 77005, USA}

\newcommand{\RomaS}{Dipartimento di Fisica, Universit\`{a} La Sapienza, P. le A. Moro 2, Roma, Italy}

\newcommand{\SCIPP}{University of California at Santa Cruz, Santa Cruz, CA 95064}
\newcommand{\Sejong}{Department of Physics and Astronomy, Sejong University, Seoul, 143-747, Korea}

\newcommand{\SHAO}{Shanghai Astronomical Observatory (SHAO), Nandan Road 80, Shanghai 200030, China}

\newcommand{\SISSA}{SISSA - International School for Advanced Studies, Via Bonomea 265, 34136 Trieste, Italy}
\newcommand{\SLAC}{SLAC National Accelerator Laboratory, Menlo Park, CA 94025}
\newcommand{\SMU}{Southern Methodist University, Dallas, TX 75275}

\newcommand{\Stanford}{Stanford University, Stanford, CA 94305}
\newcommand{\StonyBrook}{Stony Brook University, Stony Brook, NY 11794}
\newcommand{\STSCI}{Space Telescope Science Institute, Baltimore, MD 21218}

\newcommand{\Syracuse}{Syracuse University, Syracuse, NY 13244}

\newcommand{\UAS}{Department of Astronomy/Steward Observatory, University of Arizona, Tucson, AZ  85721}
\newcommand{\UAM}{Universidad Aut\'onoma de Madrid, 28049, Madrid, Spain}

\newcommand{\UCBP}{Department of Physics, University of California Berkeley, Berkeley, CA 94720, USA}
\newcommand{\UCD}{University of California at Davis, Davis, CA    95616}
\newcommand{\UCI}{University of California, Irvine, CA 92697}

\newcommand{\UCL}{University College London, WC1E 6BT London, United Kingdom}
\newcommand{\UCR}{University of California at Riverside, Riverside, CA 92521}

\newcommand{\UCSC}{University of California at Santa Cruz, Santa Cruz, CA 95064}
\newcommand{\UCSD}{University of California San Diego, La Jolla, CA 92093}
\newcommand{\UFL}{University of Florida, Gainesville, FL 32611}

\newcommand{\UGTO}{Divisi\'on de Ciencias e Ingenier\'ias, Universidad de Guanajuato, Le\'on 37150, M\'exico}

\newcommand{\Umich}{University of Michigan, Ann Arbor, MI 48109}
\newcommand{\UMN}{University of Minnesota, Minneapolis, MN 55455}

\newcommand{\UNIPD}{Dipartimento di Fisica e Astronomia ``G. Galilei'',Universit\`a degli Studi di Padova, via Marzolo 8, I-35131, Padova, Italy}
\newcommand{\UNM}{University of New Mexico, Albuquerque, NM 87131}

\newcommand{\UoM}{Jodrell Bank Center for Astrophysics, School of Physics and Astronomy, University of Manchester, Oxford Road, Manchester, M13 9PL, UK}

\newcommand{\UR}{Department of Physics and Astronomy, University of Rochester, 500 Joseph C. Wilson Boulevard, Rochester, NY 14627, USA}
\newcommand{\UrbanaC}{Department of Physics, University of Illinois at Urbana-Champaign, Urbana, Illinois 61801, USA}

\newcommand{\UTD}{University of Texas at Dallas, Texas 75080}

\newcommand{\UWaterloo}{Department of Physics and Astronomy, University of Waterloo, 200 University Ave W, Waterloo, ON N2L 3G1, Canada}
\newcommand{\UWMadison}{Department of Physics, University of Wisconsin - Madison, Madison, WI 53706}
\newcommand{\UW}{University of Washington, Seattle 98195}

\newcommand{\VSI}{Van Swinderen Institute for Particle Physics and Gravity, University of Groningen, Nijenborgh 4, 9747~AG~Groningen, The~Netherlands}
\newcommand{\VT}{Virginia Tech, Blacksburg, VA 24061}

\newcommand{\WCA}{Centre for Astrophysics, University of Waterloo, Waterloo, Ontario N2L 3G1, Canada}

\newcommand{\Wyoming}{Department of Physics and Astronomy, University of Wyoming, Laramie, WY 82071, USA}
\newcommand{\Yale}{Department of Physics, Yale University, New Haven, CT 06520}

\newcommand{\sumnu}{\Sigma m_\nu}
\newcommand{\eV}{\mathrm{eV}}

\begin{document}
\date{}

\title{\bf Astro2020 Science White Paper\\\vspace{0.2in} Neutrino Mass from Cosmology: \\Probing Physics Beyond the Standard Model}
\maketitle
\noindent \textbf{Thematic Areas:}  Cosmology and Fundamental Physics \\

\noindent\textbf{Principal Author:}
\\
Name: Cora Dvorkin\\
Institution: Department of Physics, Harvard University, Cambridge, MA 02138, USA\\
Email: \texttt{cdvorkin@g.harvard.edu}\\
Phone: (617)-384-9487\\

\begin{center}
\textbf{Authors:}
  \linebreak
{Cora Dvorkin$^{1}$, 
Martina Gerbino$^{2}$, 
David Alonso$^{3}$, 
Nicholas Battaglia$^{4}$, 
Simeon Bird$^{5}$, 
Ana Diaz Rivero$^{1}$, 
Andreu Font-Ribera$^{6}$, 
George Fuller$^{7}$, 
Massimiliano Lattanzi$^{8}$, 
Marilena Loverde$^{9}$
}
\end{center}
\begin{center}
{
% ...
}\\[0mm]
 \end{center}
 
\noindent
{\scriptsize 
$^{1}$ \HarvardPhys \\
$^{2}$ \ANLHEP \\
$^{3}$ \Oxford \\
$^{4}$ \Cornell \\
$^{5}$ \UCR \\
$^{6}$ \UCL \\
$^{7}$ \UCSD \\
$^{8}$ \INFNFerrara \\
$^{9}$ \StonyBrook \\
% ...
}

\normalsize
\pagebreak

\clearpage
\begin{abstract}
Recent advances in cosmic observations have brought us to the verge of discovery of the absolute scale of neutrino masses. Nonzero neutrino masses are known evidence of new physics beyond the Standard Model. Our understanding of the clustering of matter in the presence of massive neutrinos has significantly improved over the past decade, yielding cosmological constraints that are tighter than any laboratory experiment, and which will improve significantly over the next decade, resulting in a
guaranteed detection of the absolute neutrino mass scale.
\end{abstract}

\thispagestyle{empty}

\newpage \setcounter{page}{1}
\section{Introduction}

In the Standard Model (SM) of particle physics neutrinos are expected to be massless, as it is not possible to build a neutrino mass term given the symmetries and the particle content of the SM. 
Nonetheless, the observed flavor oscillations in solar and atmospheric neutrinos are only possible if neutrinos are massive, representing a striking evidence for physics beyond the SM (BSM). It is therefore clear that understanding the value of neutrino masses is one of the key questions in fundamental physics.

From a theoretical standpoint, there are two main avenues to give the neutrinos mass. 
Adding right-handed neutrino fields in a minimal extension of the SM can generate a \textit{Dirac} mass term $m_D$ for neutrinos through their coupling to the Higgs boson field. 
In such a scheme, 
the smallness of neutrino masses, with respect to the charged fermions that acquire mass through the same mechanism, is puzzling in itself.
If neutrinos were \textit{Majorana} particles, it is also possible to write Majorana mass terms  
generated by some unknown physics at a high-energy scale $m_M$ much above the electroweak scale. 
The interplay between the Dirac and Majorana mass terms makes the neutrino ``split'' into a heavy component with mass $m_\mathrm{heavy} \simeq m_M$ and a light component
with mass $m_\mathrm{light} \simeq m_D^2/m_M \ll m_D$. This is the well-known see-saw mechanism \cite{Giunti:2007ry}:
the higher the scale $m_M$ is pushed, the lower the mass of the light neutrino state becomes.

Neutrino oscillation experiments can measure two of the neutrino-mass splittings \cite{deSalas:2017kay}, and are getting very close to a determination of the neutrino-mass ordering (see preliminary results from T2K collaboration\footnote{\url{https://zenodo.org/record/1286752}} and e.g. \cite{Abi:2018dnh} for future prospects). However, they have no information about the absolute scale of the neutrino masses, $\sumnu$. Cosmology, on the other hand, is a promising avenue for the determination of $\sumnu$. 
Massive neutrinos leave unique imprints on cosmological observables throughout the history of our universe \cite{Lesgourgues:2006nd,Wong:2011ip,Lesgourgues:2012uu,PhysRevD.98.030001,Lattanzi:2017ubx}. 
Current cosmological observations already provide the tightest bounds on the sum of the neutrino masses \cite{Aghanim:2018eyx}, although they are unable to go beyond a very tight upper limit. As next-generation surveys approach, their improved sensitivity will help reach a guaranteed target for physics beyond the SM. Cosmology is likely to be the first experimental avenue to move from a tight upper limit to a clear detection of $\sumnu$. Experimental efforts are also being devoted to a first direct detection of the cosmic neutrino background (e.g. the PTOLEMY experiment \cite{Betts:2013uya}), which represents a very challenging task.

Note that cosmological observables are not the only probes of the absolute neutrino mass scale. Complementary information can be provided by laboratory searches such as kinematic measurements in $\beta$-decay experiments \cite{Drexlin:2013lha} and neutrino-less double-$\beta$ decay ($0\nu2\beta$) searches \cite{Cremonesi:2013vla,DellOro:2016tmg}. A detection of the absolute neutrino mass scale with cosmology would be crucial to test the consistency between different probes. In fact, an inconsistent picture would be an interesting indication of new physics in the neutrino sector.

The aim of this white paper is to highlight how cosmology can help shed light on the still-unknown value of the neutrino masses. In Section \ref{sec:growth_structure} we briefly review the effect of massive neutrinos on the growth of structure in the universe, and we outline different cosmological probes that can be used to improve our knowledge of the absolute neutrino mass scale. In Section \ref{sec:optical_depth} we quote the sensitivity to $\sumnu$ in light of expected improvements on some limiting factors such as uncertainties in the optical depth to reionization as well as theoretical uncertainties in the dark energy equation of state. In Section \ref{sec:synergy} we discuss the synergy between cosmology and laboratory searches as a tool for improving our understanding of BSM physics, and we make our concluding remarks in Section \ref{sec: conclusions}.

\section{Cosmological probes of massive neutrinos}
\label{sec:growth_structure}

In addition to contributing to the expansion history of the universe through their energy density, a more peculiar imprint of massive neutrinos is that they alter the evolution of matter perturbations.
A meaningful physical scale to define is the free-streaming scale, $k_\mathrm{fs}=0.018\,\Omega_m^{1/2}\left[m_\nu/(1\,\eV)\right]\,h\,\mathrm{Mpc^{-1}}$,
roughly corresponding to the size of the particle horizon at the time of the neutrino non-relativistic transition. At scales $k \gg k_\mathrm{fs}$, neutrinos exhibit large thermal velocities and do not contribute to the clustering of structures, while at scales $k \ll k_\mathrm{fs}$, neutrinos effectively behave as a cold dark matter (CDM) component. 
Thus, the growth of matter perturbations at small scales gets delayed, as perturbations evolve in a mixed matter-radiation environment rather than the purely matter-dominated environment at large scales.

An outline of different cosmological probes that can potentially be used to improve our constraints on the sum of neutrino masses in the next decade is layed out below.\\

\noindent {\underline{\bf CMB Lensing}}:
The large-scale structure (LSS) of the universe deflects the path of cosmic microwave background (CMB) photons traveling from the last-scattering surface to Earth. The deflection angle is, to leading order, the gradient of the lensing potential, and the lensing power spectrum is proportional to the integrated distribution of matter along the line of sight. CMB lensing thus probes the matter directly on nearly-linear scales, and has the benefit that the source (the CMB) is very well understood. 
Furthermore, low levels of foreground systematics are expected when using polarization lensing reconstruction. 
Larger neutrino masses imply a larger neutrino energy density and less clustering on small scales, therefore the overall effect of massive neutrinos is a reduction of the lensing power at intermediate and small scales \cite{Kaplinghat:2003bh}. 
\\

\noindent {\underline{\bf Galaxy Clustering}}:
Galaxies reside in the gravitational potentials of dark matter halos, tracing the overall structures of the universe, and their distribution is therefore affected by the presence of massive neutrinos \cite{Hu:1997mj}. 
Linear redshift-space distortions in the clustering of spectroscopic galaxy surveys can be used to measure the amplitude of density fluctuations at low redshift \cite{Kaiser1987}.
In combination with a prior on the amplitude of scalar fluctuations ($A_s$) from CMB experiments, future spectroscopic surveys can provide one of the tightest constraints on the sum of neutrino masses \cite{Font-Ribera2014}.

Massive neutrinos have a second effect: on very large, linear scales, the galaxy power spectrum has a step-like feature corresponding to the free-streaming length of neutrinos.
In addition to the suppression of the matter power spectrum, neutrinos produce a scale-dependent galaxy bias due to their free-streaming nature, which partially compensates the suppression due to neutrino mass~\cite{Villaescusa-Navarro:2013pva,Castorina:2013wga,LoVerde:2014pxa,Munoz:2018ajr,Chiang:2017vuk,Chiang:2018laa}. To fully take advantage of next-generation surveys, we must improve our modelling of the effect of massive neutrinos on non-linear scales, and N-body simulations will be required (see \cite{Brandbyge:2008js, Viel:2010bn,2012MNRAS.420.2551B,2013MNRAS.428.3375A,Banerjee:2016zaa,Bird:2018all,Valcin:2019fxe} for different attempts).
\\

\noindent {\underline{\bf Optical Lensing}}: 
Tomographic weak lensing measurements from photometric redshift surveys will provide a direct probe of the growth of structure as a function of time. This is complementary to both galaxy clustering and CMB lensing, and is a vital observable in order to disentangle the effects of a non-zero neutrino mass from those of, for example, non-standard dark energy scenarios \cite{2018PhRvD..97l3544M}.\\

\noindent {\underline{\bf Galaxy-Lensing Cross-Correlation}}:
The cross-correlation between the lensing power spectrum from next-generation CMB experiments (such as the Simons Observatory \cite{Ade:2018sbj} or CMB-S4 \cite{Abazajian:2016yjj}) and the galaxy power spectrum from future galaxy surveys is a promising handle on $\sumnu$, since both probes are sensitive to the amplitude of matter fluctuations. Their cross-correlation has the ability to reduce effects from systematic contamination affecting each probe individually. 
\\

\noindent {\underline{\bf Sunyaev-Zel'dovich Cluster Abundances}}:
Next-generation CMB experiments will provide extended catalogues of clusters detected through the thermal Sunyaev-Zel'dovich (tSZ) signal. The abundance of clusters as a function of their mass and redshift is a proxy for the evolution of structures and, therefore, can provide useful insights on $\sumnu$. A major source of uncertainty is the cluster mass calibration. However, with future surveys, two independent pathways for calibration will be available: internally via CMB halo lensing or externally via optical weak lensing. The  higher redshift sources, from e.g. WFIRST, will be important for calibration.
Although clusters are complex systems, if systematic uncertainties can be reduced, they represent an independent avenue to tight constraints on $\sumnu$. Most of their power sits in the redshift dependence, which is potentially able to reduce the physical degeneracy between $\sumnu$ and dark energy parameters \cite{Madhavacheril:2017onh}. \\

\noindent {\underline{\bf Kinetic Sunyaev-Zel'dovich}}:
Next-generation CMB experiments will also provide high signal-to-noise measurements of the kinematic Sunyaev-Zel'dovich (kSZ) effect. This effect is proportional to the integrated momentum along the line of sight of free electrons with respect to the CMB rest frame. Thus, kSZ measurements constitute a new powerful probes of the peculiar velocity distribution of clusters. Velocities probe the cosmological growth rate, which can constrain the sum of the neutrino masses, among other extensions to $\Lambda$CDM \cite{Mueller2015b}. Currently, the major source of systematic uncertainty is the degeneracy of this effect with the optical depth of galaxies or clusters \cite{Battaglia2016,ALBF2016,SmithKSZ2018}.\\ 

\noindent {\underline{\bf Lyman-$\alpha$ forest}}:
As the light from distant quasars travels towards us, it is incrementally affected by the absorption of intergalactic hydrogen, a tracer of the underlying density. 
This phenomenon, known as the Lyman-$\alpha$ forest, is a unique probe of the growth of structure on small scales, covering a redshift range ($2 < z < 5$) that is inaccessible by current galaxy surveys. The combination of this measurement with the amplitude of CMB fluctuations provides one of the tightest constraints on $\sum m_\nu$ \cite{Pal-Del2015}, which is expected to further improve with future surveys such as DESI \cite{Font-Ribera2014}.\\

\noindent {\underline{\bf Cosmic Voids}}:
The large free-streaming length of neutrinos prevents their clustering within dark matter halos and galaxies \cite{Paco_2011, Paco_2012, Ichiki_Takada, LoVerde:2014rxa}, and it also inhibits the evacuation of neutrinos from cosmic voids. Thus, while non-linear evolution will empty voids of CDM and baryons, neutrinos will barely feel the voids dynamic. For this reason, voids are probably the only environment where the fraction of neutrinos over CDM $+$ baryons can be much larger than the cosmological fraction, boosting the amplitude of the effect of neutrinos with respect to other cosmological observables \cite{PisaniMassaraSpergel_2019}. 
The statistical properties of voids, as identified in both the Lyman-$\alpha$ forest \cite{Paco_2011} or galaxy surveys \cite{Massara_2015,Banerjee:2016zaa,Kreisch_2018},
can be used to break the degeneracy between $\sum m_\nu$ and $\sigma_8$ (the amplitude of matter fluctuations on $8h^{−1}$ Mpc scales), which limits the amount of information that can be extracted from standard probes, such as galaxy clustering.

\section{Sensitivity to $\sumnu$ and parameter degeneracy}
\label{sec:optical_depth}
Many of the observables mentioned in the previous section depend on a measurement of the amplitude of primordial fluctuations from the CMB, which is limited by our knowledge of the reionization optical depth $\tau$. 
When the first galaxies reionize the intergalactic medium, a new source of polarization pattern in the CMB arises due to scattering of CMB photons off free electrons.  
The new scattering induces an overall power suppression proportional to $e^{-2\tau}$ in CMB spectra at intermediate and small scales. 
This suppression affects cosmological constraints on $\sumnu$, as it limits our ability to compare the amplitude of primordial fluctuations from the CMB to the amplitude of matter perturbations from late-universe probes. Therefore, in the absence of probes that break the degeneracy between the amplitude of matter perturbations and the neutrino mass, a better determination of $\tau$ is a key target for the next decade.

CMB constraints of $\tau$ can be obtained from improved measurements of large-angular-scale ($\ell < 30$) CMB E-modes. Several experimental efforts are devoted to this goal (CLASS \cite{2014SPIE.9153E..1IE}, BFORE \cite{Bryan:2017sdk,Bryan:2018jgc}, LiteBIRD \cite{Matsumura:2013aja} and PICO \cite{2019arXiv190210541H}).
Measurements of the 21-cm signal, such as those from HERA \cite{Liu:2015txa}, can also provide a better determination of $\tau$. 
This type of measurements are technologically challenging and come with the difficulties of having to separate the faint 21-cm signal from the much brighter foreground contamination from our galaxy. 
Another avenue to improve constraints on $\tau$ is to use the small-scale kSZ effect from reionization. By optimally combining the information in the kSZ 4-point function, the reionization and late-time parts
of the signal could be isolated \cite{Ferraro:2018izc}.

With the current sensitivity of $\sigma(\tau) = 0.007$ \cite{Aghanim:2018eyx}, next-generation surveys will result in an almost $3\sigma$ detection of the minimal mass scenario allowed by oscillation experiments. 
An optimal combination of next-generation CMB and LSS surveys has the potential to reach a sensitivity of $\sigma(\sumnu)\sim 14$ meV, corresponding to a nearly-$4\sigma$ detection of the minimal mass scenario.

Another source of theoretical uncertainty in the detection of neutrino masses from cosmology is the degeneracy between $\sumnu$ and other cosmological parameters that control the evolution of the universe at late times, such as the dark energy equation-of-state parameter $w$. 
Geometrical measurements (such as Baryon Acoustic Oscillations, BAO) or tomographic measurements of the late-time universe will be sensitive to the different redshift dependence of the signatures that massive neutrinos and a non-standard dark energy component have on cosmological probes. This will partially break the degeneracy between $\sumnu$ and $w$, and will increase the robustness of neutrino mass estimates from cosmology.

\section{Synergy with laboratory searches}
\label{sec:synergy}
\begin{wrapfigure}{R}{0.6\textwidth}
   \centering
 \includegraphics[width=.45\textwidth]{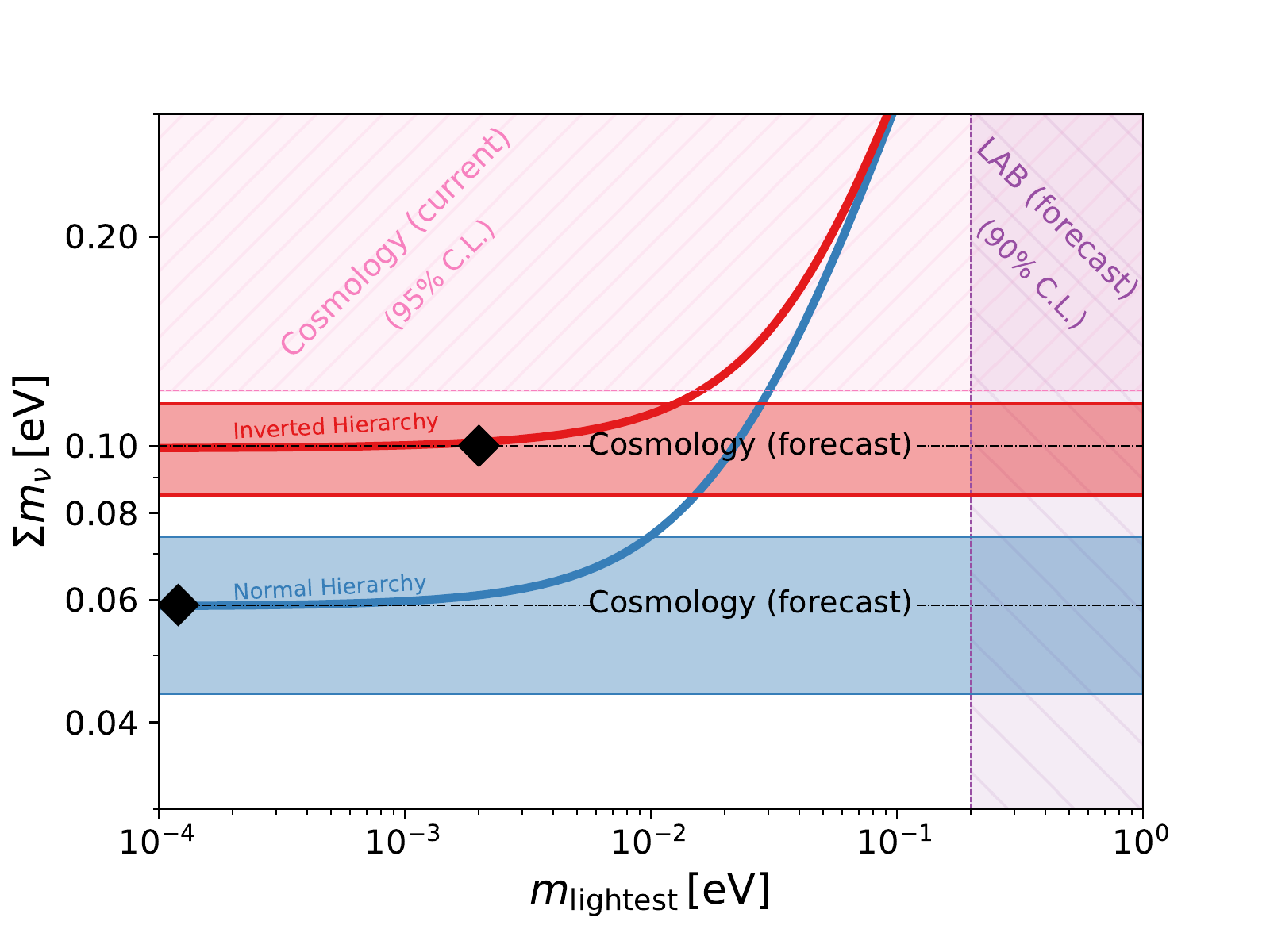}
   \caption{\footnotesize Forecasted sensitivities from future cosmological surveys and a cosmic-variance limited measurement of the optical depth to reionization are shown in horizontal bands for two cases: $\sumnu=0.06\,\eV$ and normal hierarchy (blue band), $\sumnu=0.10\,\eV$ and inverted hierarchy (red band). Current constraints from CMB + BAO exclude the pink horizontal region at 95\% C.L. \cite{Aghanim:2018eyx}. The expected 90\% C.L. limit from the $\beta$-decay experiment KATRIN \cite{Katrin} is shown as the vertical purple band. Note that here a normal hierarchy is assumed to translate the KATRIN limit on the neutrino effective mass $m_\beta$ to a limit on the lightest neutrino state $m_\mathrm{lightest}$. However, the difference with the inverted hierarchy is negligible on the scale of the plot.}
    \label{fig:mnuvsml}
\end{wrapfigure}

Cosmology and laboratory avenues are sensitive to different combinations of the individual neutrino masses and mixing parameters. Therefore, they can provide complementary information, as shown in Figure \ref{fig:mnuvsml}. 
In fact, $0\nu2\beta$ events could only happen if neutrinos were Majorana particles \cite{PhysRevD.25.2951}. 
In the context of a three-active-neutrino scenario, future $0\nu2\beta$ experiments could reach a $3\sigma$ discovery sensitivity of $0.020\,\eV$ \cite{Agostini:2017jim} and could be competitive with cosmological surveys. 
On the other hand, ongoing $\beta$-decay searches, such as KATRIN \cite{Katrin} and Project8 \cite{Project8}, are expected to reach a model-independent 
sub-eV sensitivity, with the possibility for Project8 of fully covering the parameter space allowed for inverted ordering. Finally, ongoing and future neutrino oscillation facilities are expected to reach a high statistical sensitivity to the neutrino mass ordering and CP violation phase. 

In such a context, several scenarios are possible. If all the above probes agree in their findings, a statistically strong and consistent detection of massive neutrino properties 
can be reached. 
On the other hand, perhaps more interestingly, significant tensions among the above probes could arise, which could possibly point to evidence of BSM physics.

\section{Conclusion}
\label{sec: conclusions}

This white paper briefly discusses the effect of neutrino mass on different cosmological observables, focusing on synergies between CMB and LSS. 
Significant progress has been made on these fronts, both in our theoretical understanding and in observations.
Neither CMB nor LSS observables alone can now provide a significant detection of neutrino masses, albeit together they are guaranteed a detection of the sum of neutrino masses in the next generation of experiments.
Neutrino masses are a surefire goal of upcoming cosmological surveys, which will help unveil the properties of the elusive neutrino particles in the next decade.

\clearpage
\bibliographystyle{unsrt}
\bibliography{references}

\end{document}